# Interacting topological magnons in a checkerboard ferromagnet

Heng Zhu[a], Hongchao Shi[a], Zhengguo Tang, Bing Tang[*]

*Department of Physics, Jishou University, Jishou 416000, China*

ABSTRACT

This work is devoted to studying the magnon-magnon interaction effect in a two-dimensional checkerboard ferromagnet with the Dzyaloshinskii-Moriya interaction. By means of the first-order Green's function formalism, the influence of magnon-magnon interaction on the magnon band topology is analyzed. In order to verify that the gap-closing phenomenon is a signature for the topological phase transitions of the checkerboard ferromagnet, we display that the Chern numbers of renormalized magnon bands are distinct above and below the critical temperature. Our results show that the checkerboard ferromagnet possesses two topological phases and its topological phase can be controlled either by the temperature or applied magnetic field due to magnon-magnon interactions. Interestingly, we find that the topological phase transition occurs twice with the increase of the temperature, which is different from the result of the honeycomb ferromagnet.

## 1. Introduction

Over the past two decades, more and more attention has been paid to the topological properties of the fermion energy band theory in those electronic systems, such as graphene [1,2]. Physically, these topological band structures can be detected via the Hall effect, which is in connection with the Berry curvature all over the first Brillouin zone [3]. Until now, the ideas of topological excitations have been extended to uncharged bosonic systems, such as phonons[4], polarons[5], and magnons[6]. In magnetic systems, magnons do not feel a Lorentz force, which generally induces the Hall effect of the electronic system. Interestingly, the temperature gradient can derive a thermal version of the Hall effect [7,8].

[*] Corresponding author.
E-mail addresses: bingtangphy@jsu.edu.cn
[a] These authors contributed equally to this work.

The magnonic system is viewed as one ideal platform for realizing bosonic analogs of the fermionic phase. The magnetic property can be easily controlled via the applied magnetic field, hence magnonic bands can be applied to exploring the still evolving the basic laws of the band theory. In magnetic systems, the emergence of the Dzyaloshinskii-Moriya interaction (DMI) will devastate the inversion symmetry, which can result in a nonzero Berry curvature and thermal magnon Hall effect [9-13]. Because (electrically neutral) magnons are not subjected to the Lorentz force, the DMI shall play the role of the effective magnetic field in momentum space via influencing the magnon motion in magnets, which can lead to a thermal magnon Hall effect. Experimentally, this thermal magnon Hall effect has been first detected in an insulating ferromagnet $Lu_2V_2O_7$ with pyrochlore lattice structure [10]. Subsequently, the thermal magnon Hall effect has been identified in the two-dimensional kagome magnet Cu(1-3, bdc)[14]. Furthermore, it has also been investigated in ferromagnets with other lattice structures, such as the honeycomb[15,16], Lieb[17], and checkerboard[18,19].

It is worth noting that, in ferromagnets or antiferromagnets, both the Heisenberg exchange interaction and the DMI are intrinsically nonlinear. Unfortunately, most of theoretical works on topological magnons have been restricted to the linear spin wave approximation, where the magnon-magnon interaction has been ignored. At the low temperature limit, interaction effects can be deemed to be frozen out. As the temperature increases, the effect of the magnon-magnon interaction will become more and more significant. Recently, the effect of the magnon-magnon interactions on Dirac magnons in two-dimensional honeycomb ferromagnets has been first studied by Pershoguba and his coworkers[20]. They have shown that magnon-magnon can cause the remarkable momentum-dependent renormalization of band structures. However, the DMI has been ignored in their work, though it has been experimentally identified in some materials. Afterwards, Mook *et al.* have found that the orientation of the DMI can influence the topological character and the way of the magnon-magnon interactions in the two-dimensional Heisenberg honeycomb ferromagnet[21]. Furthermore, Lu *et al.* have investigated the topological property of Dirac magnons in honeycomb ferromagnets by incorporating a second-nearest-neighbor DMI and the magnon-magnon interaction into the Hamiltonian of the system [22]. They have displayed that the magnon-magnon interactions can induces a topological phase transition of Dirac magnons, which is depicted via the sign transformation of the thermal Hall conductivity. Very recently, Sun *et al.* have shown that the honeycomb ferromagnetic spin system with a nontrivial DMI can reveal much interesting physics when the effect of magnon-magnon interaction is considered [23]. More importantly, they have realized flat bands with the nontrivial topology, which can provide one realistic platform to investigate those strongly correlated bosonic systems.

In this article, we will theoretically investigate topological magnons in a Heisenberg ferromagnet on the two-dimensional checkerboard lattice. Inspired by the ideas form the PRL [22], the effect of magnon-magnon interaction on the band topology shall be analyze by making use of Green's function formalism up to the first-order in the perturbation theory. We shall incorporate DMI and magnon-magnon interactions into our model Hamiltonian to study the topological aspect of the checkerboard ferromagnet. We show that topological phase transitions in checkerboard ferromagnets are driven by the magnon-magnon interactions and these transitions are marked via the sign variation of the thermal Hall conductivity. The topological phase of the checkerboard ferromagnet can be tuned either by the temperature or external magnetic field. The corresponding continuous topological phase transitions driven by these magnon-magnon interactions are accompanied with a gap-closing phenomenon. More details on this work will be fully displayed in the following sections.

## 2. Ferromagnetic checkerboard lattice model

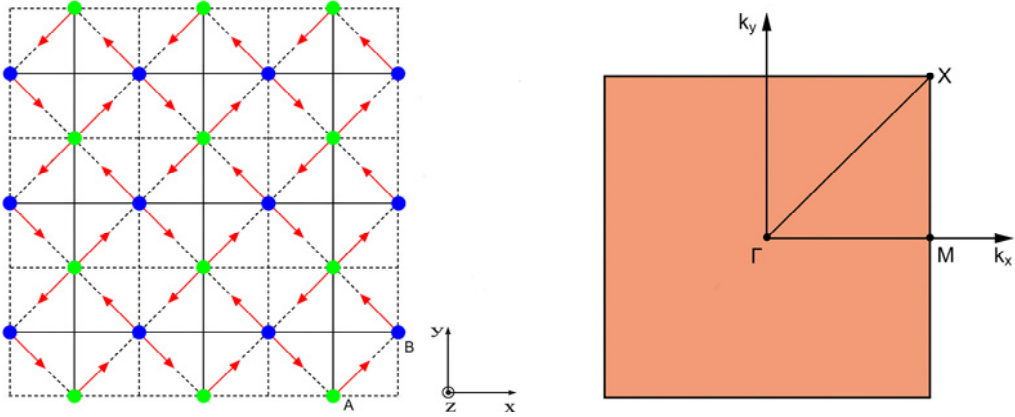

**Fig. 1.** (Color online) A scheme of 2D checkerboard lattice (left panel) and its first Brillouin zone (right panel). Both $J$ and $D$ stand for the first-nearest-neighbor interactions between localized spins at lattice sites A and B. $J'$ corresponds to the second-nearest-neighbor interactions between lattice sites A and A, and B and B, being along the solid line. In the first Brillouin zone, there exist a path marked linking the $\Gamma$, $X$, and $M$ three points, where $\Gamma = (0,0)$, $M = (\pi, 0)$, and $X = (\pi, \pi)$.

In this research, let us begin with one Heisenberg ferromagnet on the two-dimensional checkerboard lattice [19,24]. Theoretically, the checkerboard lattice can be regarded as one 2D projection of the 3D pyrochlore lattice. Some compounds (such as $Sr_2F_2Se_2$ and $Na_2Ti_2Sb_2O$) can be described via the checkerboard lattice[25], and new ferromagnetic materials with checkerboard lattice structure are expected to synthesized in the future. The Hamiltonian of our checkerboard ferromagnet is governed by

$$\mathcal{H} = -J\sum_{\langle i,j \rangle} \vec{S}_i \cdot \vec{S}_j - J'\sum_{\langle\langle i,j \rangle\rangle} \vec{S}_i \cdot \vec{S}_j + \sum_{\langle i,j \rangle} \vec{D}_{ij} \cdot (\vec{S}_i \times \vec{S}_j) - g\mu_B \vec{H} \cdot \sum_i \vec{S}_i, \qquad (1)$$

where $\vec{S}_i$ corresponds to the localized spin operators. The first two terms are respectively first-nearest-neighbor and second-nearest-neighbor ferromagnetic exchange interactions ($J>0$ and $J'>0$). The third term denotes the first-nearest-neighbor DMI, where $\vec{D}_{ij} = v_{ij} D \vec{e}_z$ is the DMI vector between lattice sites $i$ and $j$. It is noted that $v_{ij} = \pm 1$ is an orientation-dependent constant in analogy with the Kane-Mele model. The last term is the coupling to the external magnetic field $\vec{H} = H_z \vec{e}_z$. We define $h = g\mu_B H_z$, where $g$ and $\mu_B$ stand for the $g$-factor and Bohr's magneton, respectively. As displayed in Fig. 1, there exist two inequivalent lattice sites A and B in the present checkerboard spin lattice. One can see the ferromagnetic Heisenberg interaction $J$ and the DM interaction $D$ between first-nearest-neighbor magnetic ions at sites A and B. Furthermore, the ferromagnetic exchange coupling interaction $J'$ is present between second-nearest-neighbor sites A and A, and B and B.

In order to bosonize the Heisenberg Hamiltonian, we need to employ the Holstein-Primakoff (HP) transformation. Truncated to zeroth order, the HP transformation can be written as $S^z = S - a^+ a$, $S^+ = \sqrt{2S} a = (S^-)^+$, where $a^+$ ($a$) corresponds to the magnon creation (annihilation) operator, and $S^\pm = S^x \pm iS^y$ stand for the local spin raising and lowering operators. Substituting the zeroth order HP

transformation into Eq. (1), one can obtain the following bosonized Hamiltonian

$$\mathcal{H} = \omega_0 \sum_i \left(a_i^+ a_i + b_i^+ b_i\right) - J_D \sum_{\langle i,j \rangle} \left(a_i^+ b_j e^{i\varphi_{ij}} + \text{H.c}\right) - f_2 \sum_{\langle\langle i,j \rangle\rangle} \left(a_i^+ a_j + b_i^+ b_j + \text{H.c}\right) \quad (2)$$

with $\omega_0 = 4J_1 S + 2J_2 S + h$, $J_D = \sqrt{J'^2 + D^2} S$, and $f_2 = \frac{1}{2} J'S$. The phase $\varphi_{ij} = v_{ij}\varphi = v_{ij}\arctan\left(\frac{D}{J'}\right)$ corresponds to the fictitious magnetic flux in each basic square plaquette of the ferromagnetic checkerboard spin lattice. The topological features of the magnon bands can be readily captured in the momentum representation. Let $\psi_{\vec{k}} = \left(a_{\vec{k}}, b_{\vec{k}}\right)^T$ be the spinor operators in the Fourier space, where a and b represent magnon annihilation operators on the sublattices A and B, respectively. Fourier transform of the Hamiltonian (2) then reads $\mathcal{H}_0 = \sum_{\vec{k}} \psi_{\vec{k}}^+ H_0(\vec{k}) \psi_{\vec{k}}$, where

$$H_0(\vec{k}) = \begin{pmatrix} \omega_0 - 2J'Sp_{\vec{k}} & -4JS\gamma_{\vec{k}} - 4iDSm_{\vec{k}} \\ -4JS\gamma_{\vec{k}} + 4iDSm_{\vec{k}} & \omega_0 - 2J'Sp'_{\vec{k}} \end{pmatrix} \quad (3)$$

with

$$\gamma_{\vec{k}} = \frac{1}{4}\sum_{n=1}^{4} e^{i\vec{k}\cdot\vec{\delta}_n}, \quad m_{\vec{k}} = \frac{1}{4}\sum_{n=1}^{4}(-1)^n e^{i\vec{k}\cdot\vec{\delta}_n}, \quad p_{\vec{k}} = \frac{1}{2}\sum_{n=1,3} e^{i\vec{k}\cdot\vec{\zeta}_n}, \quad p'_{\vec{k}} = \frac{1}{2}\sum_{n=2,4} e^{i\vec{k}\cdot\vec{\zeta}_n}, \quad (4)$$

where $\vec{\delta}_n$ is the vectors linking the nearest neighbors and $\vec{\zeta}_n$ is the vectors linking second nearest neighbors, and $v_{ij}$ is absorbed in $(-1)^n$. The matrix $H_0(\vec{k})$ also can be written as

$$H_0(\vec{k}) = h_0(\vec{k}) \cdot \sigma_0 + h_x(\vec{k}) \cdot \sigma_x + h_y(\vec{k}) \cdot \sigma_y + h_z(\vec{k}) \cdot \sigma_z \quad (5)$$

with

$$h_0(\vec{k}) = 4JS + 2J'S + h - J'S\left(p'_{\vec{k}} + p_{\vec{k}}\right),$$

$$h_x(\vec{k}) = -4JS\gamma_{\vec{k}}, \quad h_y(\vec{k}) = 4DSm_{\vec{k}}, \quad h_z(\vec{k}) = J'S\left(p'_{\vec{k}} - p_{\vec{k}}\right),$$

(6)

where $\sigma_0$ is the identity matrices and $\sigma_x, \sigma_y$ and $\sigma_z$ are the Pauli matrices. The dispersion relations of the upper and lower bands are reads

$$\varepsilon_{\pm}^0(\vec{k}) = h_0(\vec{k}) \pm \varepsilon(\vec{k}), \quad (7)$$

where $\varepsilon(\vec{k}) = \sqrt{h_x^2(\vec{k}) + h_y^2(\vec{k}) + h_z^2(\vec{k})}$.

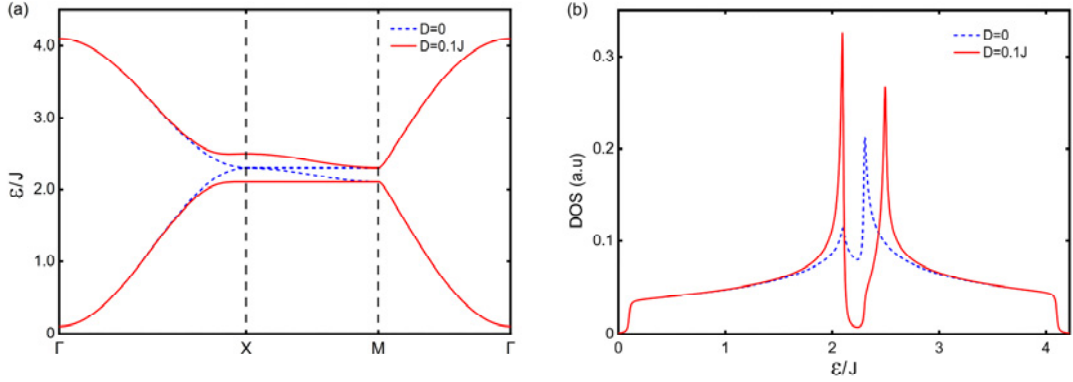

**Fig. 2.** (Color online) (a) Magnonic band structures of the checkerboard ferromagnet along the path $\Gamma - X - M - \Gamma$. (b) Density of states per unit cell of the ferromagnet on the checkerboard lattice. The parameters are chosen as $J' = 0.1J$, $h = 0.1J$, and $S = \dfrac{1}{2}$.

From Eq. (7), we note that the magnon energy spectrum of the present checkerboard ferromagnet contains two energy bands, i.e., the optical "up" and acoustic "down" bands. In Fig. 2(a), we plot magnon energy dispersion curve of the 2D checkerboard ferromagnet. In the absence of the DM interaction, i.e., $D = 0$, the "up" and "down" band touch at the X point. Physically, the DM interaction can destroy the inversion symmetry of the checkerboard ferromagnetic lattice, which opens a magnon band gap $\Delta\varepsilon = 8DS$ at the X point. Fig. 2(b) shows the density of states per unit cell as a function of energy for the two energy bands in Eq. (6). The interesting properties of the present system are manifested near the X points. When the band gap opens, the band structure is topologically nontrivial and Chern numbers of the two magnon bands are nonzero. For the checkerboard ferromagnet with the DMI, the corresponding Berry curvatures, spin Hall conductivity, spin Nernst coefficients, and magnon thermal Hall conductivity have been calculated in Refs.[18,19 ].

## 3. Interacting topological magnons

*3.1. The renormalization of magnon band*

We consider the first-order expansion of $S^{-1/2}$ in HP transformation to obtain the magnon-magnon interaction in the real lattice space, the interacting part $\mathcal{H}_{int}$ of the full Hamiltonian can be expressed as

$$\begin{aligned}
\mathcal{H}_{int} = &\frac{J}{4} \sum_{\langle i,j \rangle} \left( b_j^+ b_j^+ b_j a_i + b_j^+ a_i^+ b_j b_j + a_i^+ a_i^+ a_i b_j + a_i^+ b_j^+ a_i a_i - 4 a_i^+ b_j^+ a_i b_j \right) \\
&+ \frac{iD}{4} \sum_{\langle i,j \rangle} v_{ij} \left( b_j^+ b_j^+ b_j a_i - b_j^+ a_i^+ b_j b_j - a_i^+ a_i^+ a_i b_j + a_i^+ b_j^+ a_i a_i \right) \\
&+ \frac{1}{2} \frac{J'}{4} \sum_{\langle\langle i,j \rangle\rangle} \left[ a_j^+ a_j^+ a_j a_i + a_j^+ a_i^+ a_j a_j + a_i^+ a_i^+ a_i a_j + a_i^+ a_j^+ a_i a_i - 4 a_i^+ a_j^+ a_i a_j \right] \\
&+ \frac{1}{2} \frac{J'}{4} \sum_{\langle\langle i,j \rangle\rangle} \left[ b_j^+ b_j^+ b_j b_i + b_j^+ b_i^+ b_j b_j + b_i^+ b_i^+ b_i b_j + b_i^+ b_j^+ b_i b_i - 4 b_i^+ b_j^+ b_i b_j \right].
\end{aligned}$$

(8)

In Fourier space the Hamiltonian is given by

$$\begin{aligned}
\mathcal{H}_{int} = &\frac{J}{N} \sum_{\{\vec{k}_i\}} \left( \gamma_{\vec{k}_4}^* b_{\vec{k}_1}^+ b_{\vec{k}_2}^+ b_{\vec{k}_3} a_{\vec{k}_4} + \gamma_{\vec{k}_2} b_{\vec{k}_1}^+ a_{\vec{k}_2}^+ b_{\vec{k}_3} b_{\vec{k}_4} + \gamma_{\vec{k}_4} a_{\vec{k}_1}^+ a_{\vec{k}_2}^+ a_{\vec{k}_3} b_{\vec{k}_4} + \gamma_{\vec{k}_2}^* a_{\vec{k}_1}^+ b_{\vec{k}_2}^+ a_{\vec{k}_3} a_{\vec{k}_4} - 4 \gamma_{\vec{k}_4 - \vec{k}_2} a_{\vec{k}_1}^+ b_{\vec{k}_2}^+ a_{\vec{k}_3} b_{\vec{k}_4} \right) \\
&- i \frac{D}{N} \sum_{\{\vec{k}_i\}} \left( m_{\vec{k}_4}^* b_{\vec{k}_1}^+ b_{\vec{k}_2}^+ b_{\vec{k}_3} a_{\vec{k}_4} - m_{\vec{k}_2} b_{\vec{k}_1}^+ a_{\vec{k}_2}^+ b_{\vec{k}_3} b_{\vec{k}_4} - m_{\vec{k}_4} a_{\vec{k}_1}^+ a_{\vec{k}_2}^+ a_{\vec{k}_3} b_{\vec{k}_4} + m_{\vec{k}_2}^* a_{\vec{k}_1}^+ b_{\vec{k}_2}^+ a_{\vec{k}_3} a_{\vec{k}_4} \right) \\
&+ \frac{J'}{4N} \sum_{\{\vec{k}_i\}} \left( p_{\vec{k}_4}^* + p_{\vec{k}_2} + p_{\vec{k}_4} + p_{\vec{k}_2}^* - 4 p_{\vec{k}_4 - \vec{k}_2}^* \right) a_{\vec{k}_1}^+ a_{\vec{k}_2}^+ a_{\vec{k}_3} a_{\vec{k}_4} \\
&+ \frac{J'}{4N} \sum_{\{\vec{k}_i\}} \left( p_{\vec{k}_4}'^* + p_{\vec{k}_2}' + p_{\vec{k}_4}' + p_{\vec{k}_2}'^* - 4 p_{\vec{k}_4 - \vec{k}_2}'^* \right) b_{\vec{k}_1}^+ b_{\vec{k}_2}^+ b_{\vec{k}_3} b_{\vec{k}_4}.
\end{aligned}$$

(9)

The momentum is conserved in all of the above interaction terms by $\frac{1}{N} \sum_i e^{i(\vec{k}_1 + \vec{k}_2 - \vec{k}_3 - \vec{k}_4) \cdot \vec{r}_i} = \delta_{\vec{k}_1 + \vec{k}_2, \vec{k}_3 + \vec{k}_4}$. All the interaction terms can be expressed in a more compact form: $\mathcal{H}_{int} = \sum_{\{\vec{k}_i\}} V_{\vec{k}_3, \vec{k}_4}^{\vec{k}_1, \vec{k}_2} \psi_{\vec{k}_1}^+ \psi_{\vec{k}_2}^+ \psi_{\vec{k}_3} \psi_{\vec{k}_4}$, where the summation runs over all $(\vec{k}_1, \vec{k}_2, \vec{k}_3, \vec{k}_4)$ combinations constrained by the momentum conservation.

According to the perturbation technique for the single-magnon Green's function developed by Sun *et al.*[23], one can obtain a first order Dyson equation on interacting magnons, which has the following form

$$G_R(\vec{k},\vec{k}';\omega) = G_R^{(0)}(\vec{k},\vec{k}';\omega) + G_R^{(0)}(\vec{k},\vec{k}';\omega) \sum\nolimits_{H_{\text{int}}}^{(1)}(\vec{k}) G_R^{(0)}(\vec{k},\vec{k}';\omega). \tag{10}$$

Here, $G_R(\vec{k},\vec{k}';\omega)$ is called the interacting Green's function, $G_R^{(0)}(\vec{k},\vec{k}';\omega) = \dfrac{\delta_{\vec{k},\vec{k}'}}{\omega - H_0(\vec{k})}$ is known as the free magnon Green's function, and $\sum\nolimits_{H_{\text{int}}}^{(1)}(\vec{k})$ stands for the first-order self-energy.

By expanding $\left[\psi_{\vec{k}}, \mathcal{H}_{\text{int}}\right]$, one can get the Hartree-type self-energy, namely,

$$\sum\nolimits_{H_{\text{int}}}^{(1)}(\vec{k}) = \begin{pmatrix} \Sigma_{11} & \Sigma_{12} \\ \Sigma_{21} & \Sigma_{22} \end{pmatrix} \tag{11}$$

with

$$\Sigma_{11} = \frac{2J}{N} \sum_{\vec{q}} \left( \gamma_{\vec{q}} \cos\phi_{\vec{q}} \sqrt{1-\chi_{\vec{q}}^2} \mu_{\vec{q}}^- - \gamma_0 \mu_{\vec{q}}^+ - \gamma_0 \sum_{\vec{q}} \chi_{\vec{q}} \mu_{\vec{q}}^- \right) + \frac{2D}{N} \sum_{\vec{q}} m_{\vec{q}} \sin\phi_{\vec{q}} \sqrt{1-\chi_{\vec{q}}^2} \mu_{\vec{q}}^-$$
$$+ \frac{J'}{N} \sum_{\vec{q}} \left( p_{\vec{q}} + p_{\vec{k}} - p_{\vec{k}-\vec{q}} - p_0 \right) \left( \mu_{\vec{q}}^+ - \chi_{\vec{q}} \mu_{\vec{q}}^- \right),$$

$$\Sigma_{12} = -\frac{2J}{N} \sum_{\vec{q}} \left( \gamma_{\vec{k}-\vec{q}} e^{i\phi_{\vec{q}}} \sqrt{1-\chi_{\vec{q}}^2} \mu_{\vec{q}}^- - \gamma_{\vec{k}} \mu_{\vec{q}}^+ \right) + i\frac{2Dm_{\vec{k}}}{N} \sum_{\vec{q}} \mu_{\vec{q}}^+,$$

$$\Sigma_{21} = -\frac{2J}{N} \sum_{\vec{q}} \left( \gamma_{\vec{q}-\vec{k}} e^{-i\phi_{\vec{q}}} \sqrt{1-\chi_{\vec{q}}^2} \mu_{\vec{q}}^- - \gamma_{\vec{k}} \mu_{\vec{q}}^+ \right) - i\frac{2Dm_{\vec{k}}}{N} \sum_{\vec{q}} \mu_{\vec{q}}^+,$$

$$\Sigma_{22} = \frac{2J}{N} \sum_{\vec{q}} \left( \gamma_{\vec{q}} \cos\phi_{\vec{q}} \sqrt{1-\chi_{\vec{q}}^2} \mu_{\vec{q}}^- - \gamma_0 \mu_{\vec{q}}^+ + \gamma_0 \sum_{\vec{q}} \chi_{\vec{q}} \mu_{\vec{q}}^- \right) + \frac{2D}{N} \sum_{\vec{q}} m_{\vec{q}} \sin\phi_{\vec{q}} \sqrt{1-\chi_{\vec{q}}^2} \mu_{\vec{q}}^-$$
$$+ \frac{J'}{N} \sum_{\vec{q}} \left( p'_{\vec{q}} + p'_{\vec{k}} - p'_{\vec{k}-\vec{q}} - p'_0 \right) \left( \mu_{\vec{q}}^+ + \chi_{\vec{q}} \mu_{\vec{q}}^- \right),$$

$$\tag{12}$$

where

$$\mu_{\vec{q}}^+ = f\left(\varepsilon_-^0(\vec{q})\right) + f\left(\varepsilon_+^0(\vec{q})\right), \quad \mu_{\vec{q}}^- = f\left(\varepsilon_-^0(\vec{q})\right) - f\left(\varepsilon_+^0(\vec{q})\right). \tag{13}$$

Here, $f\left(\varepsilon_\pm^0(\vec{q})\right) = \left[e^{\beta \varepsilon_\pm^0(\vec{q})} - 1\right]^{-1}$ is the Bose-Einstein distribution function, $\beta = 1/k_B T$, $\phi_{\vec{q}} = \arg\left[-h_x(\vec{q}) + i h_y(\vec{q})\right]$, and $\chi_{\vec{q}} = \dfrac{h_z(\vec{q})}{h(\vec{q})}$.

In the low temperature, the second-order effect for the DMI can be neglected [23]. Here, following Ref. [22], we will only consider the first-order renormalized Hamiltonian. It is not difficult to obtain the renormalized Hamiltonian for our system, which is

$$H_1(\vec{k}) = H_0(\vec{k}) + \sum\nolimits_{H_{\text{int}}}^{(1)}(\vec{k})$$
$$= h'_0(\vec{k})\sigma_0 + h'_x(\vec{k})\sigma_x + h'_y(\vec{k})\sigma_y + h'_z(\vec{k})\sigma_z,$$

(14)

with

$$h'_0(\vec{k}) = \omega_0 + v_0 - (J'S - Q'_0)p'_k - (J'S - Q_0)p_k \quad, \quad h'_x(\vec{k}) = -(4JS - M_x)\gamma_{\vec{k}} \quad,$$

$$h'_y(\vec{k}) = (4DS - M_y)m_{\vec{k}} \quad, \quad h'_z(\vec{k}) = z_0 + (J'S - c'_0)p'_k - (J'S - c_0)p_k \quad,$$

(15)

where

$$v_0 = \frac{2J}{N}\sum_{\vec{q}}\left(\gamma_{\vec{q}}\cos\phi_{\vec{q}}\sqrt{1-\chi_{\vec{q}}^2}\mu_{\vec{q}}^- - \gamma_0\mu_{\vec{q}}^+\right) + \frac{2D}{N}\sum_{\vec{q}}m_{\vec{q}}\sin\phi_{\vec{q}}\sqrt{1-\chi_{\vec{q}}^2}\mu_{\vec{q}}^-$$
$$- \frac{J'}{2N}\sum_{\vec{q}}(p_{\vec{q}} - p'_{\vec{q}})\chi_{\vec{q}}\mu_{\vec{q}}^- + \frac{J'}{2N}\sum_{\vec{q}}(p_{\vec{q}} + p'_{\vec{q}} - 2p_0)\mu_{\vec{q}}^+,$$

$$Q_0 = \frac{J'}{2N}\sum_{\vec{q}}(1 - p_{\vec{q}})(\mu_{\vec{q}}^+ - \chi_{\vec{q}}\mu_{\vec{q}}^-), Q'_0 = \frac{J'}{2N}\sum_{\vec{q}}(1 - p'_{\vec{q}})(\mu_{\vec{q}}^+ + \chi_{\vec{q}}\mu_{\vec{q}}^-),$$

$$M_x = \frac{2}{N}\sum_{\vec{q}}\left(J\mu_{\vec{q}}^+ - J\gamma_{\vec{q}}\cos\phi_{\vec{q}}\sqrt{1-\chi_{\vec{q}}^2}\mu_{\vec{q}}^-\right), M_y = \frac{2}{N}\sum_{\vec{q}}\left(D\mu_{\vec{q}}^+ - Jm_{\vec{q}}\sin\phi_{\vec{q}}\sqrt{1-\chi_{\vec{q}}^2}\mu_{\vec{q}}^-\right),$$

$$z_0 = -\frac{J'}{N}\sum_{\vec{q}}(p_{\vec{q}} + p'_{\vec{q}} - 2p_0)\chi_{\vec{q}}\mu_{\vec{q}}^- + \frac{J'}{N}\sum_{\vec{q}}(p_{\vec{q}} - p'_{\vec{q}})\mu_{\vec{q}}^+ - \frac{2J\gamma_0}{N}\sum_{\vec{q}}\chi_{\vec{q}}\mu_{\vec{q}}^-,$$

$$c_0 = \frac{J'}{2N}\sum_{\vec{q}}(1 - p_{\vec{q}})(\mu_{\vec{q}}^+ - \chi_{\vec{q}}\mu_{\vec{q}}^-), c'_0 = \frac{J'}{2N}\sum_{\vec{q}}(1 - p'_{\vec{q}})(\mu_{\vec{q}}^+ + \chi_{\vec{q}}\mu_{\vec{q}}^-).$$

(16)

By diagonalizing the renormalized Hamiltonian (14), one can obtain a renormalized magnon dispersion relation, which reads

$$\varepsilon_{\pm} = h'_0(\vec{k}) \pm \varepsilon'(\vec{k}), \tag{17}$$

where $\varepsilon'(k) = \sqrt{h_x'^2(k) + h_y'^2(k) + h_z'^2(k)}$.

When the temperature is very low, the interaction between two magnons can be neglected, and at the same time the self-energy effect disappears. However, as the temperature grows, the interaction between two magnons can not be ignored, and then the self-energy has to be taken into account. In fact, the renormalized magnon band can be experimentally detected via some well-established approaches, e.g., the inelastic neutron scattering, the inelastic X-ray scattering, the Brillouin light scattering and so on [26-31].

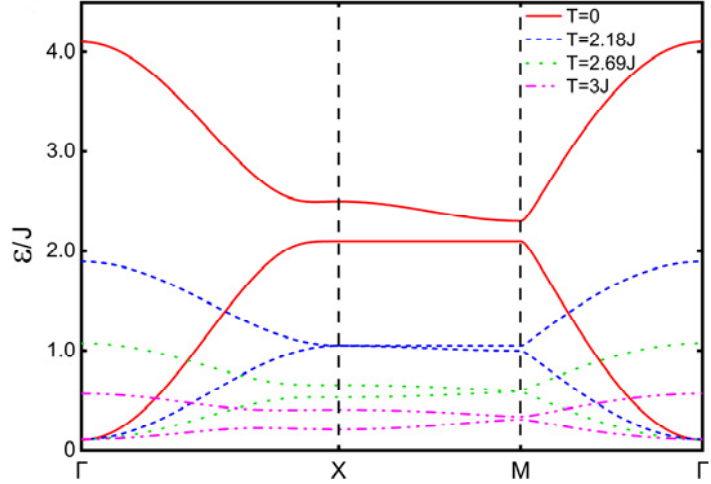

**Fig. 3.** (Color online) The renormalized magnon band structures for four different temperatures along the path $\Gamma - X - M - \Gamma$. The parameters are chosen as $J' = 0.1J$, $D = 0.1J$, $h = 0.1J$, and $S = \frac{1}{2}$.

In Fig. 3, we display the renormalized magnon band structures of the system for different value of the temperature. It is clearly seen that, with the increase of the temperature, the renormalized magnon band gap decreases and the spectrum becomes gapless at the point X decreases and will close at $T \approx 2.18J$ for this particular $h$. If we further increase the temperature, the gap reopens and its size increases with the temperature. Moreover, when $T \approx 2.69J$, the gap-closing phenomenon also occurs at the point M for this particular $h$. In order to understand the effect of the magnetic field strength on sizes of the gap, we plot the gaps at the points X and M as a function of temperature for different values of $h$, as shown in Fig. 4. One can clearly see that the critical temperature ($T_{X,c}$ for the point X and $T_{M,c}$ for the point M) for the gap-closing increases as increasing the value of $h$. For the same $h$, $T_{M,c}$ is greater than $T_{X,c}$.

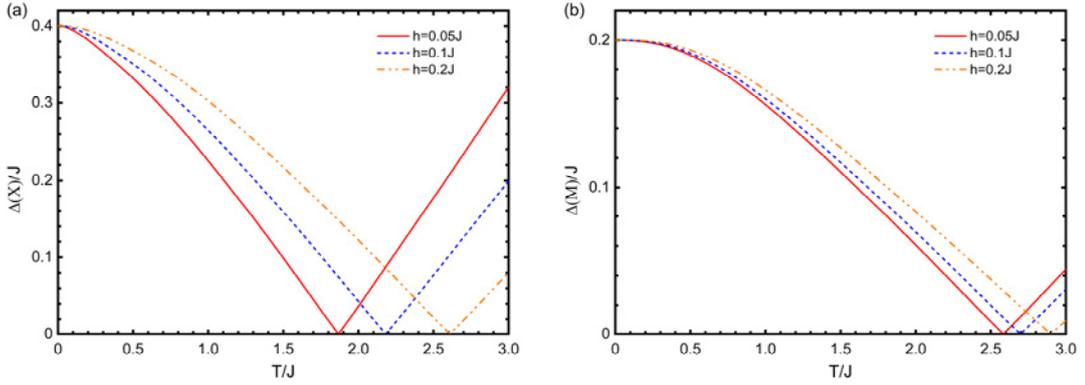

**Fig. 4.** (Color online) The gaps as a function of temperature: (a) at the point X; (b) at the point M. The parameters are chosen as $J' = 0.1J$, $D = 0.1J$, and $S = \frac{1}{2}$.

*3.2. Topological properties of interacting magnon*

In bosonic systems, the occurrence of the nontrivial band topology is caused by the energy band structure. This nontrivial band topology can be characterized via a nonzero Berry curvature, which produces a quantized integer, i.e., the Chern number.

Physically, a nontrivial band topology can arise only when the present system reveals a nontrivial gap between two magnon energy bands and each band possesses the nonzero Chern number. In 2D bosonic systems, the Berry curvature can be expressed as

$$\Omega_{\pm}(\vec{k}) = \mp \frac{1}{2h'^3} \vec{h}' \cdot \left( \frac{\partial \vec{h}'}{\partial k_x} \times \frac{\partial \vec{h}'}{\partial k_y} \right). \tag{18}$$

where, $\vec{h}' = \left( h'_x(\vec{k}), h'_y(\vec{k}), h'_z(\vec{k}) \right)$, and $h' = \sqrt{h'^2_x(\vec{k}) + h'^2_y(\vec{k}) + h'^2_z(\vec{k})}$.

The topological phase of the present system is characterized via the Chern numbers of each renormalized magnon band, which can be calculated by integrating the relevant Berry curvature over the whole first Brillouin zone (BZ), namely,

$$C_{\pm} = \int_{BZ} d^2k\, \Omega_{\pm}(\vec{k}). \tag{19}$$

Through adjusting the magnon population, we shall realize the topological phase transition in the upper magnon band from $C = 1$ to $C = -1$.

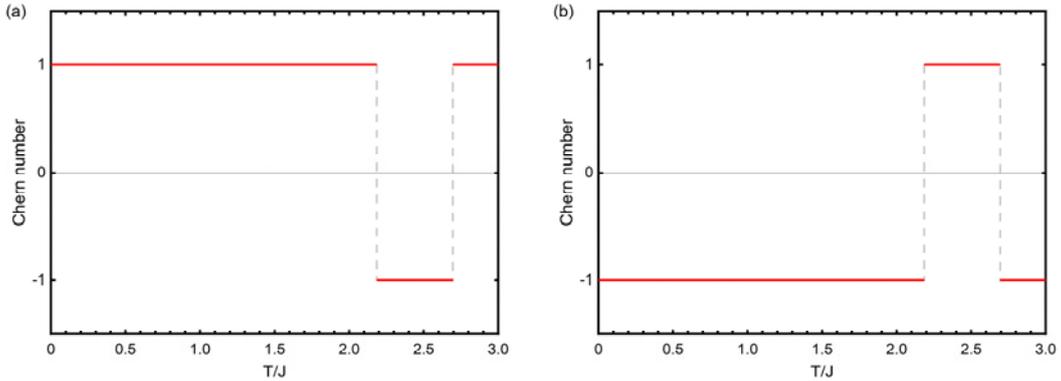

**Fig.5.** The Chern number for the renormalized magnon band versus the temperature $T$: (a) the optical "up" band, (b) the acoustic "down" band. The other parameters are chosen as $J' = 0.1J$, $D = 0.1J$, $h = 0.1J$, and $S = \frac{1}{2}$.

In Fig. 5, we display the Chern number for the renormalized magnon band as a function of the temperature $T$. It is clearly seen that the system possesses two topological phases, namely, $(C_+,C_-) = (1,-1)$ and $(C_+,C_-) = (-1,1)$. Based on the previous results, we have now realized that the renormalized magnon gap can close and reopen at the critical temperatures $T_{X,c}$ and $T_{M,c}$, which signifies that there may exist twice topological phase transition at $T_{X,c}$ and $T_{M,c}$, respectively.

For magnons in the two-dimensional ferromagnet, one longitudinal temperature gradient can give rise to a transverse heat current via the Berry curvature. Physically, such formation of the transverse heat current is referred to as the thermal Hall effect[7,32]. The thermal Hall conductivity can be expressed as

$$\kappa_{xy} = -\frac{k_B^2 T}{(2\pi)^2 \hbar} \sum_{\pm} \int d^2 k c_2(n_\pm) \Omega_\pm(\vec{k}). \tag{20}$$

Here, $n_\pm = f(\varepsilon_\pm(\vec{k})) = \left[e^{\beta \varepsilon_\pm(\vec{k})} - 1\right]^{-1}$ corresponds to the famous Bose-Einstein distribution function, $\beta = 1/k_B T$, $c_2(x) = (1+x)\left(\ln\frac{1+x}{x}\right)^2 - (\ln x)^2 - 2\text{Li}_2(-x)$, and $\text{Li}_2(x)$ represents the dilogarithm function.

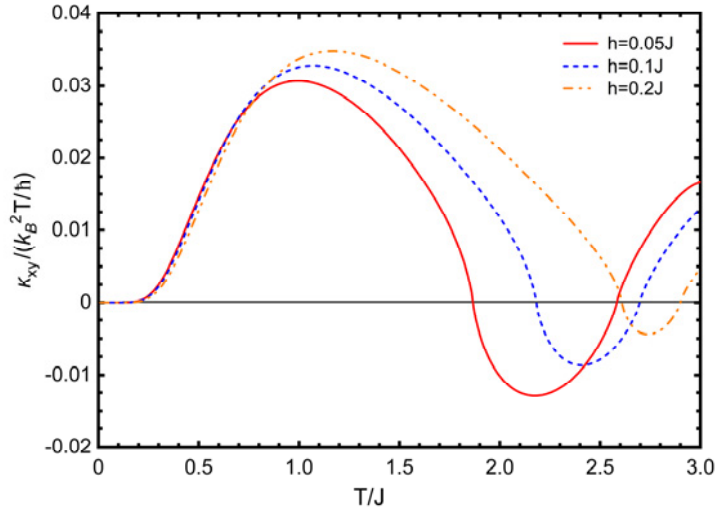

**Fig. 6.** The thermal Hall conductivity vs temperature for different values of $h$. The other parameters are chosen as $J' = 0.1J$, $D = 0.1J$, and $S = \frac{1}{2}$.

In the linear spin wave approximation, i.e, neglecting magnon-magnon interactions, $\kappa_{xy}$ is always greater than zero, as shown in Ref. [19]. When the magnon-magnon interaction is considered, the topological phase transition will occur. In Fig. 6, we show the dependence of $\kappa_{xy}$ on the temperature $T$. We can clearly see that $\kappa_{xy}$ changes continuously as the temperature increases, and is positive for $T \in [0, T_{X,c}) \cup (T_{M,c}, \infty)$ while it is negative for $T \in (T_{X,c}, T_{M,c})$. Obviously, the thermal hall conductivity and the Chern number of the upper band possess the same signs as increasing temperature. In principle, the sign reversal of the thermal Hall conductivity is a vital indicator on a topological phase transition. In Fig. 6, we find that the topological phase transition occurs twice as the temperature increases. This result is very interesting. Moreover, it is clearly seen that the critical temperature for the occurrence of the topological phase transition rise as the strength of the magnetic field increase.

## 4. Conclusions

To conclude, when incorporating the Hartree-type self-energy linked to the magnon-magnon interactions into the single-magnon Hamiltonian, the system has two topological phases: $(C_+, C_-) = (1, -1)$ and $(C_+, C_-) = (-1, 1)$. It has been shown that the topological phase can be controlled either via the temperature or applied magnetic field due to magnon-magnon interactions. Especially, we have found that the topological phase transition occurs twice with the increase of the temperature, which is a new finding. The thermal hall conductivity and the Chern number of the upper band possess the same signs as increasing temperature, thus the sign reversal of the thermal Hall conductivity can be viewed as a vital indicator on a topological phase transition of the checkerboard ferromagnet.

**Acknowledgments**

We want to thank Dr. Hao Sun for many useful suggestions. This work was supported by the National Natural Science Foundation of China under Grant No. 12064011, the Natural Science Fund Project of Hunan Province under Grant No. 2020JJ4498 and the Graduate Research Innovation Foundation of Jishou University under Grant No. Jdy21030.